\documentclass[12pt,a4paper]{article}
\usepackage{axodraw,pstricks,graphicx,epsfig,color,amssymb,amsmath,amscd}

\title{Baryogenesis by R-parity violating top quark decays
and neutron-antineutron oscillations}
\author{
A.D. Dolgov$^{\rm a,b,c}$, F.R. Urban$^{\rm a,b,d}$
\\[5mm]
${\rm ^a}$ {\small\it INFN, via Saragat 1, 44100, Ferrara, Italy} \\
${\rm ^b}$ {\small\it University of Ferrara, Department of Physics, via Saragat 1,}\\ 
{\small\it 44100, Ferrara, Italy}\\
${\rm ^c}$ {\small\it ITEP, Bol. Cheremushkinskaya 25, 117218, Moscow, Russia}\\
${\rm ^d}$ {\small\it School of Physics and Astronomy, CAPT, University on Nottingham,}\\
{\small\it University Park, Nottingham NG7 2RD, United Kingdom}
}
\date{}

\begin{document}

\newcommand{\be}{\begin{eqnarray}}
\newcommand{\ee}{\end{eqnarray}}
\newcommand{\bi}{\bibitem}
\newcommand{\lar}{\leftarrow}
\newcommand{\rar}{\rightarrow}
\newcommand{\lrar}{\leftrightarrow}
\newcommand{\mpl}{m_{Pl}}
\newcommand{\mplq}{m_{Pl}^2}
\newcommand{\rp}{${\cal R}$-parity }
\maketitle

\begin{abstract}
Generation of the cosmological baryon asymmetry in SUSY based model with broken
R-parity and low scale gravity is considered. The model allows for a long-life
time or even stable proton and observable neutron-antineutron oscillations.
\end{abstract}

\section{Introduction}
\paragraph{}
Baryogenesis is well known to be a challenging issue in nowadays cosmology and 
particle physics. We know that, at least within a confidence radius of 10 Mpc, or even
much more, see \cite{Ruj}, the universe is strongly dominated by baryons, while practically
no antibaryons are observed in the sky. The cosmological number density of baryons
with respect to the number density of photons in cosmic microwave background radiation
(CMBR) is quite small. From Big Bang Nucleosynthesis (BBN) and the angular fluctuations
of CMBR, recently accurately measured by WMAP, we know 
that (see e.g.~\cite{pdg}):
\be
\beta=\frac{n_b-n_{\bar{b}}}{n_{\gamma}}=\frac{n_B}{n_{\gamma}}\approx6\cdot10^{-10},
\label{beta}
\ee
On the other hand, this number is
not so small, as it seems, compared to a locally baryo-symmetric universe, 
where the ratio $n_b/n_\gamma = n_{\bar{b}}/{n_{\gamma}}$
would have been some 10  orders of magnitude smaller than this. 
Explaining the origin of the cosmological asymmetry between particles and antiparticles 
and the value of ratio (\ref{beta}) has proven to be an interesting meeting point 
for particle physics and cosmology, especially since, according to the seminal paper 
by Sakharov \cite{Sak}, it 
requires and gives hints about physics beyond the standard model. Indeed, 
in order for most baryogenesis models to work, baryon number nonconservation, 
C and CP violation, and departure from thermal equilibrium are required, thus pushing 
forward the demand for new fundamental physics. In fact even within the context 
of the minimal standard model (MSM) these requirements are fulfilled, but the 
magnitude of the baryon asymmetry obtained in the frameworks of MSM
is much smaller than the observed value (\ref{beta}).

There are many extensions of MSM which can explain the origin of the cosmological 
baryon asymmetry and its value, for a review see refs.~\cite{ad-bs,b-rev}. 
Doing that, one has to keep in mind a very
strong lower bound on the proton life-time~\cite{pdg}: $\tau_p > 2\cdot 10^{29}$ years
(model independent) or  $\tau_p >  10^{31}-10^{33}$ years depending upon the decay
channels. The goal of this work is to present a SUSY based model with broken
\rp which allows for a long proton life-time, or even for stable proton and
can explain the observed cosmological baryon asymmetry.
As we see in what follows, it can be done in models with low scale
gravity~\cite{low-grav}, where the short distance Planck mass $M^*$ is much smaller 
than the standard long distance Planck mass, $M_{Pl} = 1.2\cdot 10^{19}$ GeV, due to 
propagation of gravity in higher dimensions\footnote{For a string theory inspired, altough non-supersymmetric, realisation of this model see ref. \cite{koko}.}. There is also another possibility
to have a low Planck mass in the early universe and the normal large one at later
cosmological epochs, prior to primordial nucleosynthesis, due to time variation
$M^* = M^* (t) $. The latter may be induced by the coupling $R\,f\left(\phi\right)$,
where $R$ is the curvature scalar and $\phi$ is some time dependent scalar 
field~\cite{biswas}. As a by-product the suggested model may lead to potentially
observable neutron-antineutron oscillations and
B-nonconserving effects in processes with heavy quarks.

If indeed, the effective Planck mass in the early universe is about
TeV values, we would be in trouble with many versions of baryogenesis
scenarios which demand much higher energy for their realization, as 
e.g. GUT or baryo-through-lepto genesis. In this case SUSY theories
with broken \rp present a viable possibility for successful cosmological
baryogenesis and, what's more, they allow for some observable effects
in particle physics with clear signatures and possibly non-negligible
magnitudes.

In the next section the basic picture of the model is given, while the 
third section is devoted to \rp violation. 
Afterward, in the fourth section, the decay channels of heavy  particles which
could potentially lead to cosmological baryon asymmetry are considered.
In section five the out-of-equilibrium decays of massive particles in 
cosmological setup are discussed and the magnitude of baryon asymmetry
is estimated. Furthermore, the processes in particle physics with 
the variation of the baryonic charge by two units, in particular 
neutron-antineutron oscillations and processes with heavy quarks
are described (section 6).
Finally, the conclusion is devoted 
to a brief summary of the results, and some ideas for future investigations.

\section{Basic features \label{s-idea}}
\paragraph{}
In this paper we investigate the possibility of generation of the observed baryon 
asymmetry at relatively low temperatures, around or below the Electroweak (EW) phase 
transition, in the context of the Minimal Supersymmetric Standard Model (MSSM), 
extended by an addition of \rp violating term in the superpotential,
which simultaneously violates baryon-number conservation. In this section
we give a qualitative picture of how such scenario could be realised, 
while in the subsequent sections we go through technicalities and details,
in particular, we consider
a concrete SUSY model with B-nonconservation.

Normally the models of low $T$ baryogenesis demand first order phase transition 
in the primeval plasma to break thermal equilibrium because the cosmological 
expansion rate, $H=\dot a /a$, at low $T$ is
much smaller than the reactions rates. Indeed,
\be
H = \left( \frac{8\pi^3 g_*}{90}\right)^{1/2} \,\frac {T^2}{M_{Pl}} \approx
16.6 \,\left(\frac{g_*}{100}\right)^{1/2}\, \frac {T^2}{M_{Pl}},
\label{H}
\ee
where $g_*$ is the number of relativistic species contributing into 
the cosmological energy 
density; at $T\sim 100$ GeV it is about 100 in MSM and may be a factor 2 larger in MSSM.
On the other hand, the characteristic rates of reactions between elementary particles 
are:
\be
\Gamma_d \sim \alpha N m,\,\,\, {\rm and}\,\,\, \Gamma_r \sim 4\pi \alpha^2 N^2 T, 
\label{Gamma-d}
\ee
where $\Gamma_d$ is the decay rate of a gauge boson with mass $m$,
$\Gamma_r$ is the typical reaction rate, $\alpha \sim 0.01$ is the gauge
coupling constant and $N$ is the number of open channels, it could be quite
large, $N = 10-100$. 

Evidently, with $M_{Pl} \sim 10^{19}$ GeV, 
deviations from equilibrium at electroweak scale
for particles with $m\sim 100$ GeV would be about $10^{-17}$ which is surely not enough 
to create the baryonic excess (\ref{beta}). 
We see that in order for low temperature models to be efficient, it is necessary to 
require a lower Plank mass, which in turn may be provided by the same theories which 
present the necessary symmetries for keeping the proton 
stable enough, that is, extra dimensions and (super)string theories. Different
models of baryogenesis with low scale gravity
have been considered earlier in ref.~\cite{low-t-bg}.

Using eqs. (\ref{H}) and (\ref{Gamma-d}) we find that thermal equilibrium would be
strongly broken, i.e. $H/\Gamma \geq 1$, if the effective Planck mass is bounded
from above by $ M^* < 10^5 $ GeV. This estimate is obtained with $N\sim 10$
and $T\sim m \sim 100$ GeV. If we demand a weaker condition, $H/\Gamma \geq 10^{-5} $,
which may be still sufficient for generation of $\beta \sim 10^{-9}$, we would see that 
baryogenesis could be possible with $ M^* < 10^{10} $ GeV. 

There could be a complication from another side: if the Hubble parameter is too large,
then thermal equilibrium might never be created prior to baryogenesis. This 
may be even
favorable for baryogenesis, as it was argued in the case of the normal $M_{Pl}$ in
ref.~\cite{ad-al}, where baryogenesis at reheating was considered. A slow process
of preheating by the inflaton in the case of low effective Planck mass may be 
supplemented by gravitational particle production since the electroweak 
scale and the particle masses can now be close to $M^*$. Correspondingly 
heavy particles with $m\sim M^*$ could be efficiently
created. Thus we can expect the primeval plasma enriched with heavy particles
produced by gravity. An analysis of the gravitational particle production at 
the end of inflation can be found in refs.~\cite{grav-prod}.

The initial properties of the cosmological plasma depend upon the magnitude
of the Higgs field, $\phi$, during the process of (pre)heating. If $\phi$ had
already reached
its nonzero vacuum expectation value, then the produced particles would be
massive. It is possible that the plasma temperature after thermalisation
would be below the EW temperature, i.e. the plasma would be already
in EW broken phase and the EW phase transition never occurred in 
cosmological history.

In what follows we neglect possible deviations off the standard thermal evolution
and consider baryogenesis from initially equilibrium state. The nonequilibrium effects
due to a possible overabundance of heavy particles, e.g. t-quarks, would lead
to a larger baryon asymmetry.

Quite efficient breaking of thermal equilibrium in the scenario under consideration 
can be created by the following mechanism. At the temperatures above
the EW phase transition all the Standard Model (SM) particles are massless 
(except for thermal masses, which are in any case lower than the temperature at 
this stage), as well as (possibly) light SUSY particles. 
At a certain stage the phase transition takes place, and the Higgs mechanism 
gives masses to SM particles via spontaneous symmetry breaking. Then, depending upon 
the final temperature after the transition, the now massive particles could get masses 
which would be higher than the temperature after the phase transition, while some
other would remain effectively massless with $m<T$. The crucial temperature is known 
by the standard calculations to be related to the Higgs mass,
(see e.g. book~\cite{muk}). 
Given the present day lower limit on the Higgs mass, we 
see that the only SM particle which may have a mass above the final temperature after 
the EW phase transition is the top quark. We believe that SUSY partners acquire
their masses through soft symmetry breaking by the analogous Higgs effect at the 
same EW energy scale. 

If the phase transition is fast enough, so that the $t$-quark would not decay
down to the Boltzmann suppressed equilibrium density at $m_t>T$
in the course of the phase transition, then the system 
would arrive to the state when the number density of $t$-quarks at the end of
the EW phase transition would be equal to 
the number density of massless particles and grossly exceed
their equilibrium value. This makes favorable conditions for baryogenesis through
$t$-quark decays which ultimately would reduce their amount down to the
equilibrium value.

The rate of the phase transition is determined by the evolution of the temperature
dependent Higgs mass:
\be
m_\phi^2 (T) = g^2_t T^2/2 - m_0^2/2
\label{m-of-T}
\ee
where $m_0$ is the vacuum Higgs mass and $g_t \sim 1$ is the Yukawa coupling of
Higgs boson to $t$-quark. In the course of the phase transition the temperature
drops down roughly speaking by $\Delta T \sim m_0$. According to the standard law
of cosmological temperature red-shift, $\dot T = - HT$, we find that the time of 
the p.t. is about $\delta t \sim H^{-1} (\Delta T/T) \sim H^{-1} $. In realistic
case the law of temperature variation is different, due to presence of the 
time varying Higgs condensate, but not much.  On the
other hand, the decay width of $t$-quark is about
$\Gamma_t \sim \alpha\,m_t$,  (see eq. (\ref{topSM})).
Thus the decay rate would be slow in comparison with the rate of the p.t. if 
\be
M^* < 10^3 m_t \approx 10^5 \,\,{\rm GeV}
\label{t-decay}
\ee

For a bigger $M^*$ the $t$-quark decay would be faster than the rate of the p.t.
and its number density would follow the equilibrium law, $n_t \sim \exp (-m_t/T)$.
The baryogenesis through $t$-quark decays would not be efficient if $T/m_t \ll 1 $,
but if this ratio is only mildly smaller than unity then $B$-nonconserving
decays of $t$-quark might be efficient enough to create the observed $\beta$. 
Moreover, $t$-quarks could decay in the course of the p.t. creating baryon asymmetry
on the way.

Thus we see that if some of $t$-quark decay channels, or decay channels of the 
produced particles, can accommodate baryon number violation as well as CP 
nonconservation, we have fulfilled all the three basic Sakharov's requirements for a 
successful baryogenesis. In our model, allowing for B-violation coupling in the 
superpotential, we provide the possibility for the asymmetry to be generated, if at 
least one superquark is lighter than the top quark, so that some B-violating channels 
are kinematically allowed. It is also possible to account for a baryon asymmetry in
the situation in which the above-mentioned superquarks are (slightly) heavier than
the top quark, but in this case a more detailed analysis of the mechanism through
which the superpartners acquire their masses is needed.
Let us remind that baryogenesis in the minimal standard
model happens to be unsuccessful, in particular, because of very weak CP-violation
in the MSM, see e.g. the lectures~\cite{ad-varenna}. In a SUSY extended model
CP-violation is not restricted by the CKM matrix and can be much stronger. 
We neglect electroweak nonconservation of baryons induced by sphalerons because
by construction baryogenesis proceeded below the EW p.t.

In what follows we retread in detail every step outlined above. 
We first consider the presence of the baryon-number-violating term 
in the superpotential from a theoretical 
point of view, and discuss some of its phenomenological consequences. Then we will 
write down the trees of the essential processes, and estimate the 
baryon asymmetry in each case, in a purely particle physics way (no plasma, no expansion). 
Afterward, we will set these estimates in a real cosmological environment showing how the 
necessity of a lower Plank mass arises, and find that the asymmetry previously estimated 
is indeed close to the actual one. Finally, keeping in mind restrictions on the model parameters
necessary for successful generation of the baryon asymmetry, we will discuss the
predictions of the model for B-nonconserving processes in particle physics, i.e.
neutron-antineutron oscillations and heavy quark decays, and conclude.

\section{SUSY background}
\paragraph{}
The \rp violating superpotential, consistent with the gauge symmetries of the SM, as
well as with global 
supersymmetry, made up only of MSSM fields, can be written as
\be
W_{\cal R}=\mu^i L_i H_u + \frac{1}{2}\lambda^{ijk}L_i L_j E^c_k + 
\lambda'^{ijk}L_i Q_j D^c_k + \frac{1}{2}\lambda''^{ijk}U^c_i D^c_j D^c_k,
\label{Wr}
\ee
where $H_u$, $L$, $E^c$, $Q$, $U^c$, and $D^c$ are the up-type Higgs, left- 
and right-handed
lepton, left- and (up- and down-) right-handed quark chiral supermultiplets; the
superscript $c$ denotes charge conjugation, and we have suppressed all
fermion, colour and gauge indices, see e.g. \cite{OlMar,r-par}.
We notice at a glance that these couplings violate either baryon or lepton number 
separately, in particular, the first three terms proportional to $\mu$, $\lambda$ 
and $\lambda'$ have $\Delta L=1$, while the fourth with coupling $\lambda''$ has 
$\Delta B=-1$, and is the one which we will later focus our attention on.
For applications of lepton number violating terms within the context of leptogenesis see ref. \cite{far1}.
The $\lambda''$ term in the superpotential (\ref{Wr}) leads to the 
following explicit interaction Lagrangian\footnote{Note that this is not complete since 
we must include soft SUSY breaking terms consistent with the above mentioned symmetries, 
see ref. \cite{r-par}.}:
\be
{\cal L}_{int} = -\frac{1}{2}\,\lambda''^{ijk}\left(\tilde{u}^*_i\,\bar{d}_j\,d^c_k+
\tilde{d}^*_k\,\bar{u}_i\,d^c_j+\tilde{d}^*_j\,\bar{u}_i\,d^c_k\right)+h.c.
\label{L-int}
\ee
where the ${*}$ denotes the complex conjugation of a scalar field, 
and h.c. means hermitian conjugate,
and where again fermion, colour and gauge indices are suppressed.
It is essential for the future estimates that $\lambda''_{ijk}$ is antisymmetric 
with respect to the last two indices $j$ and $k$, because of the antisymmetry with respect to colour.

\rp is known to be defined as:
\be
{\cal R}=\left(-1\right)^R=\left(-1\right)^{3\left(B-L\right)+2s}
\label{Rp}
\ee
where $B$, $L$ and $s$ are baryon number, 
lepton number, and spin, respectively. It is 
straightforward to see that if \rp is exactly conserved, then every term 
in $W_{\cal R}$, eq. (\ref{Wr}),
is strictly forbidden. However, it is not necessary from a phenomenological point of 
view to forbid simultaneously all the couplings in (\ref{Wr}), since, for instance, 
the disastrously efficient tree-level diagrams for proton decay are not generated
if only the $\lambda''$ term is present. Moreover in such a theory proton must be
absolutely stable. Indeed in this case the lepton number is strictly conserved and
proton simply does not have any channel to decay into\footnote{In higher orders 
of perturbation 
theory proton may decay due to a possible small Majorana mass of neutrino or due to
exchange of the heavy Majorana fermion responsible for see-saw mechanism, if they
exist.}. On the other hand, neutron-antineutron oscillations and other processes with
variation of baryonic number by two, e.g. nuclei decays, may become observable with
a mild increase of the present-day experimental accuracy. If in addition to $\lambda''$ 
other couplings in eqs. (\ref{Wr},\ref{L-int}) are also non-vanishing they would
induce jointly proton decay but one can always take them small enough to avoid
conflicts with experimental bounds.

The discussion of all the different 
combinations of vertices, with their experimental consequences as well as the parities 
associated to them and the underlying theories which provide these symmetries, is beyond 
the scope of this paper, and we address the interested reader to the comprehensive 
review \cite{r-par}, in particular to section 2.7, references therein, and 
ref. \cite{ibque}. The main point which we mention here is that we can build a 
phenomenologically acceptable SUSY model with a superpotential which includes the 
$\lambda''$ term, and forbids all the $\Delta L\neq0$ ones. These models are usually 
embedded in (super)string theories, in which the way the extra dimensions are 
compactified, or, in the case of braneworld, the internal symmetries between the 
branes, provide the necessary symmetries for this configuration to be realised.

The next point which deserves discussion is the issue of the mass spectrum of 
the MSSM. Since 
none of the superpartners of the known SM particles are observed experimentally, 
we have only lower limits on their masses which can be found 
in ref.~\cite{pdg}. Theoretically there is no certainty about the form of the 
spectrum, but it is widely believed that one of the two mass-eigenstates of 
s-tops and/or one out of the two of s-bottoms are lighter than all the other 
s-quarks, and that some neutralinos and/or charginos are likely to be even lighter.

Lacking precise knowledge in this sector, we assume as an exemplificative toy model 
a given spectrum compatible with experimental data, and with some general requirements 
on SUSY models. To be more precise, we assume either that there is one out of the 
two mixtures of s-tops, or one of the two mixtures of the s-bottoms, which is lighter 
than the ordinary top quark, while all others s-quarks are heavier. We will specify 
later, when we consider quantitatively some concrete model, where the neutralinos 
and charginos reside in this scenario, since different possibilities are allowed and 
lead to (slightly) different results.

As a final remark let us note that if we take the lower 
limits, given by the Particle Data Group~\cite{pdg},
for the superpartner masses\footnote{Actually these limits have been found 
assuming \rp conservation, but we believe that possible \rp non-conserving decays
do not change these bounds significantly.},
whatever couple of (different) particles we choose, their total mass, $m_1+m_2$, 
is either very close to or mostly above, the top quark mass. 
Actually, the only combinations with $m_1+m_2 <m_t$
we can find are $\tilde\chi^0_1$, or $\tilde\chi^0_2$, 
and one light s-quark or chargino\footnote{Throughout the paper we denote
super-quarks as s-quarks, which should not be confused with the strange MSM quark.}.
Thus we should not consider decay channels of top quark and daughter particles where
more than one s-particle are produced, with the exceptions mentioned above, since it
appears unlikely that these channel are kinematically allowed. 

\section{Particle physics models and possible decay channels \label{s-decays}}
\paragraph{}
We outline here four different possibilities of heavy particle decay channels
determined by a different choice of their mass relations. 
Below we draw a schematic tree for the processes in every situation, 
and compute the baryon-antibaryon asymmetry generated by these decays
assuming freely decaying particles, thus neglecting 
the universe expansion and the thermal bath in which particles are immersed, where, 
given  the latter, we are neglecting inverse reaction too. At this stage we assume,
moreover, that SUSY particles are produced only by decays of heavy
SM particles and their daughter processes. In the realistic cosmological situation
when the $t$-quark became out of equilibrium after the
EW p.t. because of $m_t>T$, as is
described in sec.~\ref{s-idea}, the inverse decays induced by the superpartners 
already present in the plasma would not be essential because of the Boltzmann 
barrier and thus the assumption of absence of s-particles in the initial state 
is justified. 

To be more specific, after the EW p.t. there would be heavy t-quarks with $m_t>T$ 
but with a non-suppressed number density. There could be some other massive 
SUSY partners as well, also with $m_s>T$. If $m_s > m_t$, 
then their decays could contribute
to the baryon asymmetry. There are also some lighter particles, into which t-quark may 
decay. The inverse decay and resonance reactions 
with these particles may diminish the asymmetry. However,
since light particles thermalise quickly,  
their number density would be close to that of massless particles, i.e. $n\sim T^3$,
and their spectrum would be the usual equilibrium one with this $T$. 
If this is the case, though it is not necessary so, 
then the inverse decay and resonance scattering of light particles 
would not be important, because
the probabilities of these processes are small, $\sim \exp(-m_t/T)$, while the number 
density of t-quarks is much larger than $(m_t T)^{3/2}\exp(-m_t/T)$ 
Hence, for every decay process, we neglect additional (inverse and resonance scattering) 
reactions while considering the cosmological setup.
A more accurate study of the complete kinetics of B-nonconservation in the primeval 
plasma may possibly change 
the simple estimates presented below but not by a large factor.

In the next section we estimate the baryon asymmetry generated by t-quark decays
and show that the results are essentially the
same as presented in subsections \ref{ss-l-stop} and \ref{ss-l-s-bottom}, 
once we set the Plank mass in a certain low-scale range.

\subsection{Lightest s-top \label{ss-l-stop}}
\begin{figure}
\centering
\includegraphics[width=1\textwidth]{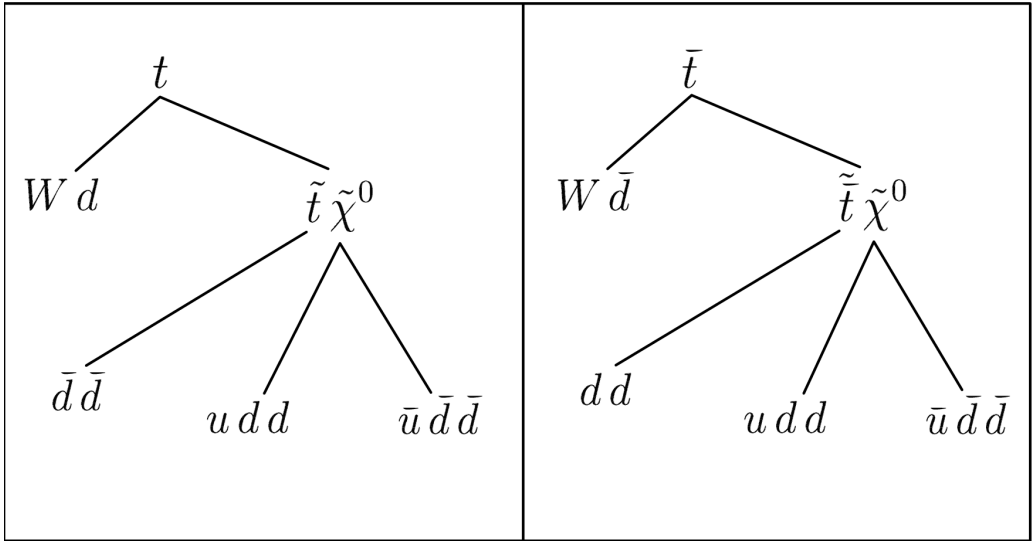}
\caption{\footnotesize{Schematic representation of the processes involved. As stated in footnote 
\ref{names}, `u' and 'd' stand for generic up- and down-type quarks, while `t' and 'b' are the 
third family top and bottom quarks. Note however that not every combination of quarks is allowed, 
see the discussion in section \ref{s-cosmo}.}}
\label{case1}
\end{figure}

\paragraph{}
Let us first consider the scenario in which the only SUSY particles lighter than the top 
quark are the s-top, $\tilde{t}_1$, and the lightest neutralino, $\tilde\chi^0_1$, where the
subscript means that we are dealing with only one mixture (labeled 1) of a certain kind of
s-particles, which is supposed to be the lightest. We begin 
with an equal initial number of top quarks and antiquarks alone, and study the evolution 
of this system. If $m_{\tilde{t}_1}+m_{\tilde\chi^0_1}<m_t$, 
then the possible processes' tree is shown in fig. \ref{case1}.

We parametrise the branching ratios in the following way\footnote{Here and in what 
follows `u' and 'd', together with their antiparticles and superpartners, mean up- and 
down-type quarks of arbitrary family, respectively, while `t' and `b' stand for top and
bottom quarks of the third family; again the definition extends to antiparticles and
superpartners.\label{names}}:
\be
\Gamma\left(t\rar W^{+}+d\right)/{\Gamma^{tot}_t}&=&1-x \nonumber\\
\Gamma\left(\bar{t}\rar W^{-}+\bar{d}\right)/\,{\Gamma^{tot}_{\bar{t}}}&=&1-\bar{x} \nonumber\\
\Gamma\left(t\rar \tilde{t}_1+\tilde\chi^0_1\right)/\,{\Gamma^{tot}_t}&=&x \nonumber\\
\Gamma\left(\bar{t}\rar \tilde{\bar{t}}_1+\tilde\chi^0_1\right)/\,{\Gamma^{tot}_{\bar{t}}}
&=&\bar{x} \nonumber\\
\Gamma\left(\tilde{t}_1\rar d+d\right)/\,{\Gamma^{tot}_{\tilde{t}}}&=&1 \nonumber\\
\Gamma\left(\tilde{\bar{t}}_1\rar \bar{d}+\bar{d}\right)/\,{\Gamma^{tot}_{\tilde{\bar{t}}}}
&=&1 \nonumber\\
\Gamma\left(\tilde\chi^0_1\rar u+d+d\right)/\,{\Gamma^{tot}_{\tilde\chi}}
&=&\frac{1-\epsilon_{\tilde\chi}}{2} \nonumber\\
\Gamma\left(\tilde\chi^0_1\rar \bar{u}+\bar{d}+\bar{d}\right)/\,{\Gamma^{tot}_{\tilde\chi}}
&=&\frac{1+\epsilon_{\tilde\chi}}{2} \nonumber
\ee
The last four processes do not conserve baryonic charge and, in the absence of
\rp breaking interaction (\ref{Wr}), $\tilde t_1$ and $\tilde \chi_1$ would be
stable. The difference between $x$ and $\bar x$ is induced by CP-violation and is small
because CP-breaking can manifest itself only in higher orders of perturbation theory due to
rescattering in the final state, see e.g.~\cite{ad-varenna,ad-yz}. The same result applies
to $\epsilon_{\tilde{\chi}}$, which is non-zero but small.

Now, if we assume that these particles decay freely
and inverse processes are negligible, we find that the final asymmetry 
generated after several characteristic lifetimes of the decaying particles, is 
given by summing all the quarks and antiquarks in the final states, with 
their respective baryonic 
charges, multiplied by the branching ratio of the channel into which they have 
been produced. This gives:
\be
\Delta B=\left(\bar{x}-x\right)-\epsilon_{\tilde\chi}\left(\bar{x}+x\right)\simeq\epsilon_t
\label{bar1a}
\ee
where $\epsilon_t=\bar{x}-x$, and in the last equality we have retained only the
first order 
terms, since we expect $\bar{x}$, $x$ as well as $\epsilon_{\tilde\chi}$ to be small.

If no CP violation is allowed in the standard model particle decay, $t\rar W+q$, 
which means that $\bar{x}=x$, we obtain:
\be
\Delta B=-2\,x\,\epsilon_{\tilde\chi},
\label{bar1b}
\ee
that is, the asymmetry is entirely generated in the second step\footnote{It is easy to see 
that even if we have more than one channel with $\Delta B\neq0$, as in fact we have, no 
asymmetry develops in the primary decay of the top quark, since all these B-violating 
channels have the same $\Delta B$.}, of the CP-violating decays of neutralinos.

Concluding this subsection we notice that if the 
condition $m_{\tilde{t}_1}+m_{\tilde\chi^0_1}<m_t$ 
is no longer satisfied because of a heavier either neutralino or s-top, 
then the baryon asymmetry cannot be generated in this version
of the scenario.

\subsection{Lightest s-bottom \label{ss-l-s-bottom}} 
\begin{figure}
\centering
\includegraphics[width=1\textwidth]{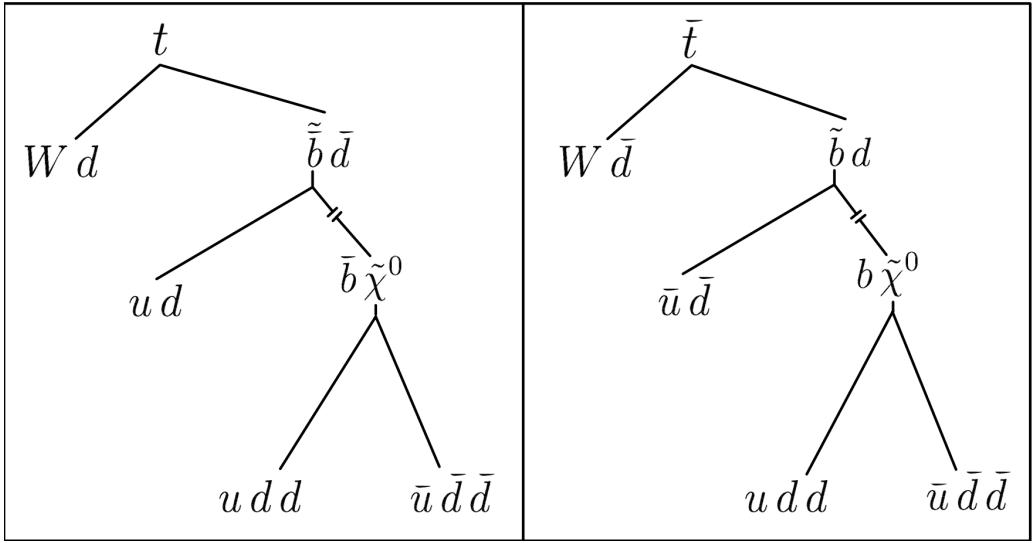}
\caption{\footnotesize{Schematic representation of the processes involved. As stated in 
footnote \ref{names}, `u' and 'd' stand for generic up- and down-type quarks, while `t' 
and 'b' are the third family top and bottom quarks. Note however that not every combination 
of quarks is allowed, see the discussion in section \ref{s-cosmo}. The line going from $\tilde{b}$ to $b\,\tilde{\chi_0}$
is crossed to indicate that this channel may be closed if the mass difference does 
not allow it.}}
\label{case2}
\end{figure}

\paragraph{}
We turn now to the case where the lightest s-quark is the s-bottom $\tilde{b}_1$, and the 
mass spectrum goes as: 
\be
m_{\tilde\chi^0_1}<m_{\tilde{b}_1}<m_t<m_{\tilde{t}_1}
\label{mass-relation}
\ee
Let us assume again that there are some
equal initial numbers of top and antitop quarks and no other particles are present.
In this case the baryon asymmetry can be generated in three steps by the decay
of t-quarks and their decay products', as sketched in fig. \ref{case2}.
The last step is actually realised only if the chain of inequalities (\ref{mass-relation})
is true all the way down. In fact, if the neutralino is heavier than the ${\tilde{b}_1}$,
the tree of decays changes, because the decay channel 
$\tilde{\bar{b}}_1\rar \bar{b}+\tilde\chi^0_1$ 
and its charge conjugated one are no longer kinematically allowed, and then this process, 
as well as 
the subsequent neutralino decay, must be cut from the tree, leaving only the 
$\Delta B\neq0$ channel. 

Then, as in the previous case, the branching ratios can be written as follows:
\be
\Gamma\left(t\rar W^{+}+d\right)/\,{\Gamma^{tot}_t}&=&1-x \nonumber\\
\Gamma\left(\bar{t}\rar W^{-}+\bar{d}\right)/\,{\Gamma^{tot}_{\bar{t}}}&=&1-\bar{x} \nonumber\\
\Gamma\left(t\rar \tilde{\bar{b}}_1+\bar{d}\right)/\,{\Gamma^{tot}_t}&=&x \nonumber\\
\Gamma\left(\bar{t}\rar \tilde{b}_1+d\right)/\,{\Gamma^{tot}_{\bar{t}}}&=&\bar{x} \nonumber\\
\Gamma\left(\tilde{\bar{b}}_1\rar \bar{b}+\tilde\chi^0_1\right)/\,
{\Gamma^{tot}_{\tilde{\bar{b}}}}&=&1-y \nonumber\\
\Gamma\left(\tilde{b}_1\rar b+\tilde\chi^0_1\right)/\,{\Gamma^{tot}_{\tilde{b}}}
&=&1-\bar{y} \nonumber\\
\Gamma\left(\tilde{\bar{b}}_1\rar u+d\right)/\,{\Gamma^{tot}_{\tilde{\bar{b}}}}
&=&y \nonumber\\
\Gamma\left(\tilde{b}_1\rar \bar{u}+\bar{d}\right)/\,{\Gamma^{tot}_{\tilde{b}}}
&=&\bar{y} \nonumber\\
\Gamma\left(\tilde\chi^0_1\rar u+d+d\right)/\,{\Gamma^{tot}_{\tilde\chi}}
&=&\frac{1-\epsilon_{\tilde\chi}}{2} \nonumber\\
\Gamma\left(\tilde\chi^0_1\rar \bar{u}+\bar{d}+\bar{d}\right)/\,{\Gamma^{tot}_{\tilde\chi}}
&=&\frac{1+\epsilon_{\tilde\chi}}{2} \nonumber
\ee
Here again the last four processes violate baryonic number conservation but in addition
the processes 3 and 4 also break it. We see that, with respect to the previously considered 
scenario, here all the three steps can generate an asymmetry, since all of them include two 
different decay channels with different $\Delta B$, either $\Delta B_{ch.1}=0$ and $\Delta B_{ch.2}=\pm1$ 
(first and second steps), or $\Delta B_{ch.1}=1$  and $\Delta B_{ch.2}=-1$ (third step).

We have chosen this parametrisation since we expect the 
ratio between $\tilde{b}_1\rar \bar{u}+\bar{d}$ and $\tilde{b}_1\rar b+\tilde\chi^0_1$ 
to be small\footnote{The estimates of the decay 
rates and the discussion on that will be given in the next section.}, 
which in turn means that $\bar{y}$ and $y$ are small, and this gives us a 
convenient way to express the baryon asymmetry generated as:
\be
\Delta B=\left(\bar{x}-x\right)-\left(\bar{x}\,\bar{y}-x\,y\right)-
\epsilon_{\tilde\chi}\left(\bar{x}+x-\left(\bar{x}\,\bar{y}+x\,y\right)\right)\simeq\epsilon_t,
\label{bar2a}
\ee
where again we could focus on the case $\bar{x}=x$ to obtain:
\be
\Delta B=-x\left(\left(\bar{y}-y\right)+\epsilon_{\tilde\chi}\left(\bar{y}+
y-2\right)\right)\simeq-x\left(\epsilon_{\tilde{b}}-2\,\epsilon_{\tilde\chi}\right)
\label{bar2b}
\ee
with $\epsilon_{\tilde{b}}$ being $\bar{y}-y$, and, as in the previous subsection, 
the last equalities in (\ref{bar2a}) and (\ref{bar2b}) follow,
once one discards higher-order terms, given the smallness of 
$\bar{x}$, $x$, $\bar{y}$, $y$ and $\epsilon_{\tilde\chi}$.

Actually, as we see in what follows, it is not mandatory that 
$\bar{y}$ and $y$ are small. Indeed, if this is not
the case, we might want to change slightly our parametrisation of the branching ratios, 
and redefine $\bar{y}$ and $y$ as $1-\bar{z}$, $1-z$, respectively, where now $\bar{z}$ 
and $z$ are small, and this gives the following result:
\be
\Delta B=\left(\bar{x}\,\bar{z}-x\,z\right)-
\epsilon_{\tilde\chi}\left(\bar{x}\,\bar{z}+x\,z\right)
\label{bar2c}
\ee
or, with no CP violation in the SM decays, given $\bar{z}-z=\epsilon_{\tilde{b}'}
=-\epsilon_{\tilde{b}}$:
\be
\Delta B=x\,\left(\bar{z}-z\right)-x\,\epsilon_{\tilde\chi}\left(\bar{z}+
z\right)\simeq x\,\epsilon_{\tilde{b}'}
\label{bar2d}
\ee

Note that if instead the case of a heavy neutralino is considered, we must set 
$y=\bar{y}=1$, or equivalently $z=\bar{z}=0$. Doing so we find that a net asymmetry 
cannot develop, thus if these scenarios are to work we must demand (at least) one
light neutralino.
As a final comment concluding this subsection we point out that there are no
kinematically allowed processes involving charginos.

\section{Cosmological setting \label{s-cosmo}}
\paragraph{}
In this section we analyse the evolution of an initial equal number density of 
top and antitop quarks in cosmological plasma. We begin by discussing initial 
conditions, then we estimate the decay rates for every process, and 
study the cosmological evolution of the baryon asymmetry. 
Since the two possibilities coming from different mass spectra described above have 
similar outcome, we will focus only on the second scenario, namely 
$m_{\tilde\chi^0_1}<m_{\tilde{b}_1}<m_t<m_{\tilde{t}_1}$.

As described in section \ref{s-idea}, before the EW phase transition all 
SM particles were massless, except for small thermal masses, since the symmetry was 
unbroken. Following the standard calculations, as given for instance in ref. \cite{muk}, 
we find the relation which links the characteristic temperature after the transition 
with the Higgs mass 
\be
T_{p.t.}^2\simeq1.5\left(m_H^2+\left(44\textrm{GeV}\right)^2\right)
\label{T-pt}
\ee
This relation in fact should be corrected in the context of MSSM, because possibly also SUSY particles would take part in the EW transition, but the effects are not very strong \cite{giu}.

Since we require the temperature after the transition to be lower than the top quark mass, 
$m_t\approx175$GeV, we have the upper bound on the Higgs boson mass $m_H\leq135$GeV, 
while with the current experimental lower bound $m_H>114.4$GeV, we find a temperature 
around 150GeV. Within this range of Higgs masses, after the EW phase transition the 
top quarks and antiquarks are the only SM particles 
which are truly non-relativistic (or better to say, semi-relativistic), thus they 
are interesting candidates for the ``parents'' of
out-of-equilibrium decay. However we could relax this 
hypothesis allowing for a wider range of Higgs masses, but in that case deviations 
from equilibrium are smaller. Nevertheless a reasonable net asymmetry will develop 
even in this case. 

We proceed through the detailed calculations of the requirements for this model to work, 
completing the discussion given in section \ref{s-idea}.
In order to know if the reactions which we are interested in, are in thermal equilibrium 
or not, we must estimate their decay rates and compare them with the expansion rate of 
the universe, given by the Hubble parameter $H=\dot{a}/a$, when $T\approx m$, where 
$m$ is the mass of the decaying particle. If $\Gamma/H\gg1$, then equilibrium would be 
established, while if 
\be
\Gamma/H\ll1
\label{req1}
\ee
the decay would be "frozen". 

The second point we must look at is the stability of top quarks and antiquarks, since 
we don't want these particles to decay before the EW transition. This is because 
otherwise, once the transition takes place, all the $t$ and $\bar{t}$, now massive, 
would decay almost instantaneously, and, since they are non-relativistic, their number 
density would be exponentially suppressed. Hence, no asymmetry will arise. This 
requirement can be written as
\be
\Gamma_{tot}\,t_{p.t.}\ll1
\label{req2}
\ee
that is, the characteristic time of the EW phase transition must be much less than 
the mean life of top and antitop quarks.

We note here that the above stated requirements of out-of-equilibrium decay and long 
lived particle, in standard cosmology result almost the in same numerical constraints, 
since $H=1/(2\,t)$ or $H=2/(3\,t)$ in radiation or dust dominated universes, 
respectively. Then, when we fix the temperature $T$ we have also fixed the Hubble 
parameter $H$ and the time $t$ at the, almost, same numerical value. However this is 
no longer true if we consider e.g. brane worlds or loop quantum gravity cosmologies, 
and, without going that far, even an universe dominated by an inflaton in its 
oscillating regime. In non standard cosmologies $H^2\neq\rho$ with $\rho$ being 
the energy density of the dominant component which fills the universe, 
generally $H^2\sim L(\rho)$ and therefore $H\neq t^{-1}$, see reviews \cite{lang, bojo} 
and references therein. In the case of an oscillating 
inflaton dominated universe, we have instead that $H\sim T^4$,
which again gives a different relation between time and temperature, which however depends on the specific inflationary model, see \cite{kolb}. In both these two cases we have two different numerical constraints for our requirements (\ref{req1}) and (\ref{req2}), but in what follows we assume the normal evolution specified at the beginning of
this paragraph.

At this point we have formulated the two basic requirements on the decay rates, then 
we list below the rates we will use afterwards.
\be
\Gamma\left(t\rar W+d\right)&=&\frac{g^2}{64\pi}\frac{V_{t\,d}^2}{m_W^2}\,m_t^3 \label{topSM}\\
\Gamma\left(t\rar \tilde{\bar{b}}_1+\bar{d}\right)&=&\frac{\lambda''^2_{t\,b\,d}}
{128\pi}\,Z_{\tilde{b}_1}^2\,m_t \label{topB} \\
\Gamma\left(\tilde{\bar{b}}_1\rar \bar{b}+\tilde\chi^0_1\right)&\approx&\frac{g^2}
{10^2\pi}\,Z_{\tilde{b}_1}^2\,Z_{\tilde\chi^0_1}^2\,m_{\tilde{b}_1} 
\label{botChi}\\
\Gamma\left(\tilde{\bar{b}}_1\rar u+d\right)&=&\frac{\lambda''^2_{u\,b\,d}}
{64\pi}\,Z_{\tilde{b}_1}^2\,m_{\tilde{b}_1} \label{botB}\\
\Gamma\left(\tilde\chi^0_1\rar u+d+d\right)&\approx&\frac{g^2}{10^4\pi^3}
\frac{\lambda''^2_{u\,d\,d}\,Z_{\tilde\chi^0_1}^2}{m_{\tilde{q}}^4}\,m_{\tilde\chi^0_1}^5 
\label{chi}
\ee
where: $g$ is the EW coupling constant, $Z_i$ is a function of the i-th SUSY particle mixing parameters, 
$m_{\tilde{q}}$ is the mass scale of 
the heavy superquarks, and $\lambda''_{i\,j\,k}$ is the \rp violating coupling constant 
relative to the B-violating part of the superpotential (\ref{Wr}). As before, "u" 
and "d", mean up- and down-type quarks, respectively\footnote{Not every combination 
is allowed, however. For instance in the process $t\rar \tilde{\bar{b}}_1+\bar{d}$, 
only d=down and d=strange are allowed by the symmetries of the coupling matrices, 
once one takes into account colour.}, while t and b stand for top and bottom quarks; 
the definition extends to antiparticles and their (anti)superpartners.

Let us now turn to consideration of the demands we want to satisfy. We will use 
the following numerical values. The top quark mass is $m_t=175$GeV, the W mass is 
$m_W=80.4$GeV, the heavy s-quarks mass scale is $m_{\tilde{q}}\approx300$ GeV, 
$Z_{\tilde{b}_1}^2=Z_{\tilde\chi^0_1}^2=0.5$, though those values are unknown and could well be one order of magnitude lower. We set the masses of 
the unknown light s-particles to the values $m_{\tilde{b}_1}\approx100$ GeV and 
$m_{\tilde{\chi}^0_1}\approx50$ GeV. Furthermore, from now on we will drop the apices on
$\lambda''$, bearing in mind that it refers to the baryon number-violating term in the
superpotential.

Up to this point we have not specified the range which the B-violating $\lambda$ 
couplings could span. We have different processes in which almost every different possible 
$\lambda$ coupling is involved. More specifically, in process (\ref{topB}) we can have 
$\lambda_{3\,3\,1}$ or $\lambda_{3\,3\,2}$, the sub-numbers being the family which the 
quarks belong to, where the first number refers to the up-type quark involved, and the 
other two refer to the two down-type quarks, with $\lambda_{i\,j\,k}=-\lambda_{i\,k\,j}$. 
Then, in process (\ref{botB}) the possibilities are 
$\lambda_{1\,1\,3}$, $\lambda_{1\,2\,3}$, $\lambda_{2\,1\,3}$ or $\lambda_{2\,2\,3}$, 
while neutralino decay, (\ref{chi}), splits into six different processes determined by 
$\lambda_{1\,1\,2}$, $\lambda_{1\,1\,3}$, 
$\lambda_{1\,2\,3}$, $\lambda_{2\,1\,2}$, 
$\lambda_{2\,1\,3}$ or 
$\lambda_{2\,2\,3}$. 
Referring again to the review \cite{r-par} (sec. 6), we see that 
some of these couplings, are quite strongly constrained, e.g. $\lambda_{112}< 10^{-6}$,
$\lambda_{113}< 10^{-3}$, while the other could be well above unity.
The most stringent bounds on $\lambda_{ijk}$
are obtained from nucleus stability which is destroyed by the baryon-antibaryon
transformation. It is discussed in the next section in connection with
neutron-antineutron oscillations.

The huge dispersion in the values of the \rp-breaking couplings $\lambda_{ijk}$
may be related to the similar or larger dispersion in the Yukawa couplings of Higgs
boson to fermions.

Once the values of the couplings have been chosen, we are able to determine the ratios 
between the different channels of every decay process. In particular, we must ensure that 
the SM top decay is the dominant one with respect to the \rp violating one, and we need 
to know which one of the two s-bottom decays is the dominant one in order to choose the 
most convenient parametrisation of branching ratio as described in section \ref{s-decays}. 
Going through the calculations, we obtain for the top quark decay:
\be
\frac{\Gamma\left(t\rar \tilde{\bar{b}}_1+\bar{d}\right)}{\Gamma\left(t\rar W+d\right)}
&\approx&0.12\lambda^2 \\
\frac{\Gamma\left(\tilde{\bar{b}}_1\rar u+d\right)}
{\Gamma\left(\tilde{\bar{b}}_1\rar \bar{b}+\tilde\chi^0_1\right)}&\approx&7.5\lambda^2
\ee
Where both $\lambda_{313}$ and $\lambda_{323}$ are allowed in the first ratio, and four 
different couplings $\lambda_{113}$, $\lambda_{123}$, $\lambda_{213}$ and $\lambda_{223}$ are 
allowed in the second ratio. Almost all of them are allowed to be as large as unity.
We see that we indeed have, for our values of $\lambda$ couplings, that the B-violating 
top decay is subdominant, while the ratio between the two possible decays of the s-bottom 
could be either bigger or smaller than one for different values of $\lambda$, which means that both possibilities 
described by the equations (\ref{bar2a}) and (\ref{bar2b}), or (\ref{bar2c}) and 
(\ref{bar2d}), are actually possible.

We consider now the evolution of these processes in the expanding universe after the EW 
phase transition. We have already pointed out that the two basic requirements for the 
viability of this model are the longevity of the top quarks in the early universe, 
and the out-of-equilibrium feature of their decay. Let us fix, for illustrative purposes, 
the temperature after the EW transition has occurred to 150 GeV, then, if we want the 
top quark to decay at that temperature in a non-equilibrium regime, we must ensure:
\be
\frac{\Gamma\left(t\rar W+d\right)}{H_{T_{p.t.}}}\approx
\frac{\mpl}{2\cdot10^5\textrm{GeV}}\leq1
\label{lowMpl}
\ee
that translates into:
\be
\mpl\leq2\cdot10^5\textrm{GeV}
\label{mplValue}
\ee
where $H$ is given by $H^2=8\pi^3\,g_{*}^{1/2}\,T^2/90\,\mpl$, with $g_{*}\approx110$ 
being the number of relativistic degrees of freedom at that time.
We have substituted
$\Gamma\left(t\rar W+d\right)$ for the total the total decay width of
t-quark, since this is very accurately true, 
and we have begun from checking of inequality (\ref{lowMpl}) because this 
automatically gives to the top and antitop quarks a long-enough life, since 
in this case the condition
$t_{p.t.}\,\Gamma\left(t\rar W+d\right)\leq1/2$ holds. 
Equation (\ref{mplValue}) gives the basic requirement for this model to work, as it has
been discussed in section \ref{s-idea}.

We now can compute all the other interesting $\Gamma/H$ ratios, and consequently the 
baryon asymmetry generated by the processes we have considered so far. We know that 
we can approximate the kinetic equations for the processes of interest in different 
ways, which are selected by the value of the ratio $\Gamma/H_m$ where $H_m=H(T=m)$, 
$m$ being the mass of the decaying particle, see refs. \cite{ad-bs, kolb}. 
Given all the numerical values specified above, we obtain:
\be
\frac{\Gamma\left(t\rar W+d\right)}{H_{m_t}}&\approx&\frac{\mpl}{3\cdot10^5\textrm{GeV}} \\
\frac{\Gamma\left(t\rar \tilde{\bar{b}}_1+\bar{d}\right)}{H_{m_t}}
&\approx&\lambda^2\frac{\mpl}{2.5\cdot10^6\textrm{GeV}} \\
\frac{\Gamma\left(\tilde{\bar{b}}_1\rar \bar{b}+\tilde\chi^0_1\right)}{H_{m_{\tilde{b}}}}
&\approx&\frac{\mpl}{5\cdot10^6\textrm{GeV}} \\
\frac{\Gamma\left(\tilde{\bar{b}}_1\rar u+d\right)}{H_{m_{\tilde{b}}}}
&\approx&\lambda^2\frac{\mpl}{10^6\textrm{GeV}} \\
\frac{\Gamma\left(\tilde\chi^0_1\rar u+d+d\right)}{H_{m_{\tilde\chi}}}
&\approx&\lambda^2\frac{\mpl}{10^{12}\textrm{GeV}}\label{chi-dec}
\ee

All the above expressions show that if condition (\ref{mplValue}) is satisfied
then automatically it would set these ratios (much) below unity. 
This in turn means that the baryon 
asymmetry, up to some subsequent dilution processes through entropy production, is 
simply given by:
\be
\beta \approx10^{-2}\Delta B
\ee
where $\Delta B$ is given by the formulas of section \ref{s-decays}. Whilst every different 
specific value of the couplings $\lambda$ and the other parameters (mixings and so on) 
gives different $\Delta B$, we notice that there are several appealing possibilities 
which could easily deal with the experimentally observed value of $\beta$, without requiring 
unnatural high values of the various CP-violation parameters involved, 
$\epsilon\approx10^{-5}$, even without allowing CP violation in the SM decay $t\rar W+b$. In this connection we notice that despite the fact that up to date no Electric Dipole Moments have been detected for neutrons or electrons, thus disfavoring maximal CP violating phases in the MSSM, nevertheless the upper bounds \cite{pdg} could be satisfied also for large values of the CP violating phases, provided that some cancelation mechanisms are at work, see ref. \cite{far2} and references therein.

As a final remark in this section, let us note that the decay of the neutralinos, given the
expression (\ref{chi-dec}), would take place at temperatures far below BBN.
However, the
above mentioned decay width is no longer correct below the scale at which $\mpl$ recovers
its classical value, which must be before the nucleosynthesis, that is, according to
\cite{low-grav}, around 10 to 100 MeV. Once $\mpl$ reaches $10^{19}$ GeV the decay of the
neutralino proceeds almost instantaneously (in comparison with the cosmological expansion 
rate), thus preventing possible troubles with late-time
entropy production which could destroy the successful
BBN predictions. Then, since we don't specify how the
Planck mass goes up to its standard value or, in other words, which is the time
scale for this process, we don't know whether neutralino decays are in equilibrium or not.
However this will not change much the results of the scenarios
outlined above, since in most of them the asymmetry produced by their
decays contributes only a small fraction to the total asymmetry, 
except for the possibilities given by eq. (\ref{bar1b}) and (\ref{bar2b}).

\section{Baryon nonconservation in particle physics\label{n-barn}}
\paragraph{}
As we have mentioned above, one of the main goals of this work to present a model of
baryogenesis which is capable to explain the observed asymmetry (\ref{beta})
and simultaneously
does not contradict very strict lower bound on the proton life-time. 
As we see below in this section, this model with the chosen parameters
values leads to noticeable, but still below the existing limits,
neutron-antineutron oscillations and potentially observable nonconservation of
baryonic charge in heavy quark decays.
According to \cite{pdg} the characteristic time
of oscillations for free neutrons is restricted by
\be
\tau_{n\bar n} = 1/\delta m_{n\bar n} > 8.6\cdot 10^7\,\,{\rm sec},
\label{tau-nn-f}
\ee
while for bound neutrons the limit is slightly stronger:
\be
\tau_{n\bar n} > 1.3\cdot 10^8\,\,{\rm sec}
\label{tau-nn-b}
\ee
The latter is obtained from the observed stability of different nuclei at
the level of $10^{32}$ years. If a bound neutron can turn into antineutron 
inside a nucleus the latter would definitely explode. Non-observation of this process 
allowed to put the above quoted limit. The arguments go as follows.
The admixture of antineutron to neutron in a nuclei is equal to
\be
\sin \theta_{n\bar n} = \delta m /\Delta E
\label{theta-n}
\ee
where $\delta m$ is the amplitude of $(n-\bar n)$-transition in vacuum
and $\Delta E \approx 100$ MeV is the energy difference of antineutron and neutron
in a nuclei. 

The probability of $\bar n$ annihilation in a nucleus is equal to
\be
\Gamma_{n\bar n} \approx \sigma_{n \bar n} V N\,\sin^2 \theta_{n\bar n}
\label{gamma-n}
\ee
where $ \sigma_{n \bar n}$ is the $n \bar n$ annihilation cross section,
$V$ is relative velocity of $n$ and $\bar n$ and $N \sim m^3_\pi \sim (100\,\rm MeV)^3$
is the number density of nucleons inside a nucleus;
$\sigma_{n \bar n} V \approx 1/m_\pi^2$. 
Assuming, according to the data, that the life-time of nuclei is bounded by
$\tau (A\rar A-2) = 1/\Gamma_{n\bar n} > 10^{32} $ years we find
\be
\tau_{n\bar n} =1/\delta m > (1/\Delta E)  \sqrt{ \sigma_{n \bar n} V N}\,\,
\sqrt{10^{32}\,{\rm years}} 
\label{tau-n-bound}
\ee
Substituting here numerical values indicated above we obtain the result which
is incidentally quite close
to the limit obtained for free neutrons (\ref{tau-nn-b}).

In the model under consideration 
the simplest process which could lead to baryon-antibaryon transformation is (\ref{chi}).
On a slightly more fundamental level it is described by a tree diagram of 
$(ud)$-transition into $\tilde{\bar b}_1$. Then the latter captures another $d$-quark 
and turns into $\tilde\chi^0_1$. After that the process goes in charge conjugated order: 
$\tilde\chi^0_1$ transforms into $\tilde{ b}_1$ and $\tilde{b}_1$ goes into 
$(\bar u \bar d)$.

The amplitude of this process is
\be
A_{B\bar B} = \frac{\lambda^2_{112} g^2 Z^2_{\tilde \chi^0_1} Z^2_{\tilde b_1}}
{m^4_{\tilde b_1} m_{\tilde\chi^0_1}}\, \left(\psi C\psi \right)^3, 
\label{b-bar-b}
\ee 
where $C$ is the charge conjugation matrix and $\left(\psi C\psi \right)^3$ is the product
of 3 spinors of initial state quarks and 3 of the final state antiquarks.
The process of neutron-antineutron oscillations would be very efficient if $udd$ quarks 
(and antiquarks) belonged to the first generation. However, this is not so and the simplest 
process includes $u$ and $d$ quarks of the first generation and another $d$ quark of
the second generation, i.e. the strange quark $s$. Neutron-antineutron oscillations must
include transformation with change of strangeness by two units, $\Delta S =2$, and
may proceed with additional second order weak interactions, see ref. \cite{chakeu}.
\begin{center}
\fcolorbox{white}{white}{
  \begin{picture}(330,125) (150,-161)
    \SetWidth{0.5}
    \SetColor{Black}
    \ArrowLine(162,-53)(210,-101)
    \ArrowLine(162,-149)(210,-101)
    \ArrowLine(222,-149)(270,-101)
    \DashArrowLine(270,-101)(210,-101){10}
    \Photon(270,-101)(342,-101){6}{4}
    \Line(270,-101)(342,-101)
    \Text(234,-89)[lb]{\Large{\Black{$\tilde{s}$}}}
    \Text(210,-161)[lb]{\Large{\Black{$s$}}}
    \Text(150,-161)[lb]{\Large{\Black{$d$}}}
    \Text(366,-89)[lb]{\Large{\Black{$\tilde{\bar{s}}$}}}
    \Text(450,-161)[lb]{\Large{\Black{$\bar{d}$}}}
    \Text(390,-161)[lb]{\Large{\Black{$\bar{s}$}}}
    \ArrowLine(450,-53)(402,-101)
    \ArrowLine(450,-149)(402,-101)
    \DashArrowLine(342,-101)(402,-101){10}
    \ArrowLine(390,-149)(342,-101)
    \Text(306,-89)[lb]{\Large{\Black{$\tilde{\chi}^0$}}}
    \Text(150,-52)[lb]{\Large{\Black{$u$}}}
    \Text(450,-52)[lb]{\Large{\Black{$\bar{u}$}}}
  \end{picture}}
\begin{center}
\footnotesize{Graph 1. Feynman diagram leading to $n\bar\Xi$ transformation or to
$\Lambda\bar\Lambda$-oscillations. Here every quark is labeled with its proper name.}
\end{center}
\end{center}

We can do the same exercise, which is done above for $n\bar n$-transformation,
considering e.g. the process of neutron transformation
into $\bar \Xi$-hyperon. A stringent upper bound, $\lambda_{112}< 10^{-6}-10^{-7}$,
was obtained from nuclei stability with respect to this process in
ref.~\cite{rb-am} (see also the review~\cite{r-par}). A similar process would
create $\Lambda \bar \Lambda $-oscillations. Using the same arguments as presented above
we find that the life time of $\Lambda \bar \Lambda $-oscillations is bounded by 
approximately $ \tau_{\Lambda\bar \Lambda} >  10^6\,\,{\rm sec} $. A weaker bound in
comparison with $n-\bar n$ case is related to a larger energy difference of $\bar \Xi$
in comparison with $\bar n$ and to a smaller annihilation cross-section of
$n\bar \Xi $. 

Assuming the specified above mechanism for baryonic number nonconservation 
we can naively
estimate the time of $n\bar n$-oscillation using the following chain of processes: 
neutron transforms into $\Lambda$ by first order weak interaction, then $\Lambda$ goes
into $\bar\Lambda$ through the process described above and and $\bar \Lambda$ ``returns''
to antineutron again by weak interactions. According to that the time of
$n\bar n$-oscillations is equal to
\be
\tau_{n\bar n} \sim \tau_{\Lambda \bar \Lambda} \,
\frac{\left(m_\Lambda -m_n\right)^2}{\mu_{n\Lambda}^2}
\label{tau-n-n}
\ee
where $m_n$ and $m_\Lambda$ are the masses of neutron and $\Lambda$ respectively and
$\mu_{n\Lambda}$ is the matrix element or weak transition of neutron to $\Lambda$.
The latter can be expressed through the amplitude of the decay of 
$\Lambda\rar p \pi^-$, using PCAC and current algebra~\cite{aiv-rev}: 
\be
\mu_{n\Lambda} \approx {A_{\Lambda\rar p\pi^-}}/{m_\pi},
\label{mu-p-Lambda}
\ee
where $A_{\Lambda\rar p\pi^-}$ is the amplitude of the indicated decay of $\Lambda$,
which can be easily found from the life-time of $\Lambda$ (we neglected
$s$-$p$ wave problem because it is sufficient for an order of magnitude estimate
presented here. Thus we find:
\be
\tau_{n\bar n} \sim 10^{12}\tau_{\Lambda \bar \Lambda} 
\label{tau-n-n2}
\ee
This is by far beyond any existing and future experimental sensitivity. The weakness 
of the $(n\bar n)$-transformation is related to the necessity of the second order 
strangeness changing weak transition with $\Delta S =2$. 

One can avoid the weak processes with $\Delta S=2$ allowing diagrams with the 
exchange of superpartner of $W$-boson, wino, $\tilde W$, 
instead of $\tilde\chi_1$,
which permits the vertex $W\rar u\bar s$. Such diagrams lead to the amplitude
$(uuddds)$ 
suppressed only by the Cabibbo-like mixing $\sin \tilde\theta_c$  between wino and
strange super-quark plus u-quark. Hence the amplitude 
of $(uudddd)$-transition is suppressed only by first order weak interaction. But the
effect is still weak.
However, even this is not necessary and the weak processes with $\Delta S=1$ 
can be also avoided leading to the allowed time of 
$n\bar n$-oscillations just above the existing bounds. This makes further improvement
of the accuracy in search for this process quite promising.

The mechanism which could lead to noticeable $n\bar n$-oscillations is the following. 
Let us consider the Feynman diagram similar to Graph 1 but with substitution of
zino, $\tilde Z$, instead of neutralino, $\tilde \chi_1$. Now the superpartner of
s-quark, $\tilde s$, can transform into non-strange quark $d$ emitting $\tilde Z$.
This process was impossible with $Z$-boson due to unitarity of the mass matrix of
quarks. Because of that the transformation matrix $Z\bar q q$ was always proportional
to unity matrix both in the mass and flavour eigenstate basis. This is not necessarily
true in the case of non-diagonal transition of quark and its superpartner. Though the
mass matrices of quarks and s-quarks are unitary as well, but they can be different 
for quarks and their superpartners. This allows for the strangeness changing processes
in ``neutral current'' interaction with an exchange of zino \cite{chan}. Correspondingly the amplitude
$(uudddd) $ would be suppressed with respect to $(uuddss)$ only by a mild factor
$\sin^2 \tilde\theta_c$, where $\tilde\theta_c$ is Cabibbo-like mixing angle 
for s-quark mass matrix. In fact this mixing angle may be even not as small as the
Cabibbo angle in usual strangeness charging decays. This opens window for quite strong
$n\bar n$ mixing.

Let us briefly mention baryonic charger nonconservation in heavy quark decays
induced by interaction (\ref{Wr},\ref{L-int}). These processes are of course 
discussed in the literature, see e.g. review~\cite{r-par}. Our only input 
here is that the constants $\lambda_{ijk}$, connecting heavy $(t,b)$ generation
with lighter ones, cannot be as small as e.g. $\lambda_{112} < 10^{-6}$. 
Otherwise, the baryogenesis in this model would not be efficient enough.  
Unfortunately we cannot predict how small $\lambda_{ijk}$ are allowed
to realize still successful baryogenesis because of uncertainty in the
value of the effective Planck mass and mass spectrum of SUSY particles.
We could find the range of couplings, masses, and CP-violating phases
which are necessary for successful baryogenesis to predict possible range
of magnitudes of B-nonconservation effects in particle physics but this is a
complicated task which we will postpone.

\section{Conclusion}
\paragraph{}
We have discussed here a scenario of baryogenesis based on a SUSY model with
broken \rp in such a way that only baryonic charge is noticeably nonconserved,
while leptonic charges are either conserved or their nonconservation is very 
weak.

In the model presented here the baryon asymmetry of the universe can be 
produced with a sufficiently large magnitude to agree with observations but 
this demands
low scale gravity to ensure the necessary deviations from thermal equilibrium. 

On the other hand, proton would not decay despite nonconservation of the baryon
number because the decay demands nonconservation of leptonic charge, 
$L$, which may be 
arbitrary weak. In the spirit of the mentioned above hierarchy of baryo-nonconserving 
couplings grossly increasing from light to heavy generations we may assume
similar picture for lepton nonconserving couplings. If it is true, lepton nonconservation
would be weak in electronic and muonic sector but could be much stronger in
tauonic sector. Lepton number nonconserving decays of $\tau$ 
would be an interesting feature to search for. 
If leptonic charge is noticeably nonconserved only in $\tau$-sector,
proton decay would be suppressed, even with nonconserved leptonic 
charge. 

Still there are several B-nonconserving effects (with conserved $L$)
which may be 
observable in direct experiments. Neutron-antineutron oscillations may have the
characteristic time just above the existing upper bounds or particles
involving heavy quarks could decay into channels with $\Delta B = 2$. 
Unfortunately the predictions are not 
accurate and theoretical uncertainty may be as large as several (3-4 or more) orders of
magnitude due to poor knowledge of the coupling constants, $\lambda $, necessary
for baryogenesis.

If \rp is indeed broken in this way, then the lightest SUSY particle 
(LSP) should be generally unstable and cannot make the cosmological 
dark matter (DM). We are not completely sure if, in these frameworks,
is impossible to make sufficiently long-lived LSP with the life-time bigger
than the universe age but have not yet found rigorous arguments either way. If LSP
is not DM then other candidates as axion, warm sterile neutrinos, primordial black
holes, etc could be DM. 

There are several problems yet to study. An interesting one is an investigation of
proton decay in a model with heavy Majorana neutrino which is responsible for 
light neutrino masses. Another thing which could be done is a more accurate
study of kinetics of generation of baryon asymmetry with different choices of
the (s)particle mass spectrum. This however, is rather an academic problem because
we do not know the appropriate parameters (couplings and masses) and 
accurate calculation with unknown numbers are not particularly interesting, at
least at this stage. 

\section*{Acknowledgments}
We are grateful to Yu. Kamyshkov for discussion and suggestions. A.D. thanks G. Kane for discussion. 
F.U. is grateful to the Particle Theory Group of the University of Nottingham, for 
the kind hospitality during the last stage of this work, and thanks O. Seto for useful 
correspondence. F.U. is supported by INFN under grant n.10793/05.

\end{document}